\documentclass[conference]{IEEEtran}

\usepackage[numbers,sort]{natbib}
\usepackage{amsmath,amssymb,amsfonts}
\usepackage{algorithmic}
\usepackage{graphicx}
\usepackage{textcomp}
\usepackage{xcolor}
\usepackage{multirow}
\usepackage{tcolorbox}
\usepackage{hyperref}
\usepackage{enumitem}

\begin{document}

\title{Syntax and Stack Overflow: A methodology for extracting a corpus of syntax errors and fixes}

\author{\IEEEauthorblockN{Alexander William Wong,
Amir Salimi, Shaiful Chowdhury, and Abram Hindle}
\IEEEauthorblockA{Department of Computing Science,
University of Alberta\\
Edmonton, Alberta, Canada\\
\{alex.wong, asalimi, shaiful, abram.hindle\}@ualberta.ca}}


\maketitle

\begin{abstract}
One problem when studying how to find and fix syntax errors is how to get natural and representative examples of syntax errors.
Most syntax error datasets are not free, open, and public, or they are extracted from novice programmers and do not represent syntax errors that the general population of developers would make.
Programmers of all skill levels post questions and answers to Stack Overflow which may contain snippets of source code along with corresponding text and tags.
Many snippets do not parse, thus they are ripe for forming a corpus of syntax errors and corrections.
Our primary contribution is an approach for extracting natural syntax errors and their corresponding human made fixes to help syntax error research.

A Python abstract syntax tree parser is used to determine preliminary errors and corrections on code blocks extracted from the SOTorrent data set.
We further analyzed our code by executing the corrections in a Python interpreter.
We applied our methodology to produce a public data set of 62,965 Python Stack Overflow code snippets with corresponding tags, errors, and stack traces.
We found that errors made by Stack Overflow users do not match errors made by student developers or random mutations, implying there is a serious representativeness risk within the field.
Finally we share our dataset openly so that future researchers can re-use and extend our syntax errors and fixes.
\end{abstract}


\section{Introduction}
Syntax errors stymie novice and expert developers alike~\cite{Tabanao:2011:PAN:2016911.2016930,Denny:2012:SEE:2325296.2325318}.
Many researchers have sought to help developers recover from syntax errors by locating syntax errors~\cite{10.7287/peerj.preprints.1132v1} and/or fixing them~\cite{DBLP:conf/wcre/SantosCPHA18}.
One impediment to this kind of research is getting access to a representative corpus of syntax errors and their subsequent corrections.
Corporae are limited because syntax error commits are rare as open source developers typically do not commit syntax error ridden code to their repositories.

Many existing datasets suffer from limited access due to ethical concerns while other datasets lack representativeness as they are pulled from student developers rather than practicing software engineers~\cite{Pritchard:2015:FDE:2846680.2846681,kelley2018system}.
Some authors address the lack of access to syntax errors by synthetically creating their own syntax error datasets via mutation~\cite{Just:2014:MVS:2635868.2635929,10.7287/peerj.preprints.1132v1}.
Mutation does not fully address the representativeness of syntax errors as they can differ from the syntax errors of actual developers~\cite{Jimenez:2018:MRN:3239235.3240500}.
These shortcomings can be mitigated by using data containing \emph{natural} errors created by a more general population of developers, where natural is defined as ``product of human effort''~\cite{Hindle:2012:NS:2337223.2337322}.
Unfortunately, naturally made syntax errors are difficult to obtain as current methods require fine grain observations of programmer activity~\cite{Brown:2014:BLS:2538862.2538924}.

Thus the goal of this work is to reproducibly and openly produce a corpus of Python syntax errors that is both \emph{natural} and representative of syntax errors made by developers as a whole.
A dataset of \emph{natural} human made syntax errors and human revised fixes would enable future software engineering research.
Models tasked with source code completion or error detection could use this dataset to train and evaluate their performance on actual developer errors.
Statistical analysis on commonly made syntax errors and their corresponding fixes could offer insight into planning future software language specifications.
This dataset would enable novel approaches for measuring the similarity of source code to natural languages.
Potential new studies could explore the naturalness of syntax corrections, comparing how natural language errors are fixed to how code errors are fixed~\cite{7886923}.
Most importantly, a replicable and reproducible methodology to produce such a corpus would allow researchers to update and improve the dataset on their own and in the future.

We extend the SOTorrent dataset\footnote{SOTorrent Dataset: \href{https://zenodo.org/record/2273117}{https://zenodo.org/record/2273117}}, built from Stack Overflow, to extract syntax errors and corrections~\cite{DBLP:conf/msr/BaltesDT008}.
Our main contribution is a methodology for extraction of human made errors and their fixes.
With our methodology, it is important to understand how accurate the pairs of syntax errors and their fixes are, and what are the general characteristics of those syntax errors.
We answer the following research questions:

\begin{enumerate}[label=RQ\arabic*),itemindent=0.8em,start=0]
    \item How accurate is our approach for extracting pairs of syntax errors and their fixes?
    \item What are Python parse errors and corresponding runtime properties of their corrections in Stack Overflow?
    \item Are Python errors in Stack Overflow similar to errors made by student programmers?
    \item Are Python errors in Stack Overflow similar to generated errors created by mutating valid code?
\end{enumerate}

Our generated dataset is released to enable future work in source code naturalness research\footnote{Dataset of Python3.6 Natural Syntax Errors and Corrections: \href{https://doi.org/10.6084/m9.figshare.8244686.v1}{https://doi.org/10.6084/m9.figshare.8244686.v1}}.

\section{Related Work}
Syntactically incorrect code is artificially derivable, as formal programming languages provide grammar rules which can be referred to for correctness.
Random token level insertions, deletions, and replacements were performed to generate syntax errors from existing open source Java projects~\cite{Campbell:2014:SEJ:2597073.2597102}.
\citeauthor{10.7287/peerj.preprints.1132v1} created Python syntax errors from valid code mined from GitHub by applying mutations on tokens, characters, and lines~\cite{10.7287/peerj.preprints.1132v1}.
Although generated errors are appealing due to the availability of open source code,  \citeauthor{Just:2014:MVS:2635868.2635929} demonstrated limitations of using mutations for software testing research~\cite{Just:2014:MVS:2635868.2635929}.
Given the task of using mutants as replacements for real faults in automated software testing research, only 73\% of mutants were coupled to faults.
Furthermore, when accounting for code coverage, the mutant to fault coupling effect is small~\cite{Just:2014:MVS:2635868.2635929}.

Automated source code repair, like identifying and refactoring improper method names, also required a labeled dataset of valid and invalid source code~\cite{liu2019learning}.
Program repair is often viewed as different than syntax error correction because testing is performed which serves as a benchmark for repaired code, while syntax errors rely primarily on parseability.

Free and open datasets of naturally made errors and their fixes are more difficult to obtain.
Blackbox, a data collection project within the BlueJ Java development environment, requires manual staff contact for access to data and forbids the release of the raw dataset~\cite{Brown:2014:BLS:2538862.2538924}.
\citeauthor{Pritchard:2015:FDE:2846680.2846681} analyzed Python programs submitted to CS Circles, an online tool for learning Python~\cite{Pritchard:2015:FDE:2846680.2846681}.
\citeauthor{kelley2018system} studied Python code submitted by students in an introductory programming course at MIT~\cite{kelley2018system}.
Gathering this data without privileged access to the provided code submissions is difficult, limiting the reproducibility of their research.
Our research used Stack Overflow and is advantageous as the raw content is freely accessible to the internet, revisions and history is tracked, and contributors have a wide range of software engineering expertise and skill sets\footnote{\label{sosurvey}Stack Overflow 2019 Survey \href{https://insights.stackoverflow.com/survey/2019}{https://insights.stackoverflow.com/survey/2019}}.

\section{Methodology}
Our methodology for finding syntax error and fix pairs is: to leverage the historical view of Stack Overflow extracted from the SOTorrent project~\cite{DBLP:conf/msr/BaltesDT008}; parse detected Python source snippet histories; extract pairs of failed and fixed revisions; validate pairs; and finally record successfully evaluated pairs.

\subsection{Code Snippets from SOTorrent}
The SOTorrent schema contains a \textbf{PostBlockVersion} table which stores version history on the latest \textbf{Posts}, which must be either a question or answer~\cite{DBLP:conf/msr/BaltesDT008}.
We queried a subset of rows from the \textbf{PostBlockVersion} table where content type equaled \texttt{CodeSnippet} and the corresponding \textbf{Post} was tagged with a term matching \texttt{python}.
If the corresponding \textbf{Post} was typed \texttt{Answer}, we used tags defined in the referenced \texttt{Question}.
To limit \texttt{TabError} exceptions, extraneous white space is removed from the content while preserving nested indentation.

\subsection{Parsing Abstract Syntax Tree}
To determine the presence of syntax errors, we attempted to parse abstract syntax trees (AST) given the extracted code.

We tackled this problem using Python's built-in \texttt{ast} module, which compiles source code into an AST object.
Valid code parses without error, while invalid code raises errors such as \texttt{IndentationError}, \texttt{TabError}, \texttt{SyntaxError}, or \texttt{MemoryError}.
The specific error message, offending line number and column offset are stored for the indentation, tab, and syntax errors (memory errors offer no such metadata).

\subsection{Filtering Candidates using Block Versions}
One issue is that many extracted code snippets containing a \texttt{python} matching tag are not valid Python, but instead contain stack traces, script execution commands, tabular data, markup, etc.
Further candidate filtering is necessary to distinguish invalid Python code from unrelated text.

To address this issue, we queried for an existing prior version within the \textbf{PostBlockVersion} table for each code snippet that successfully parsed into an AST.
If the prior version contained a parse error, we stored these two versions of code as our syntax error and fix.
Unrelated snippets that are never fixed into valid Python source code are removed from analysis.

\subsection{Runtime Validation with the Interpreter}
We checked for code snippet validity by running the corrected source code.
Code that validly parsed into an AST may still be faulty due to various errors, like importing invalid modules or referencing undefined variables.
Source code that cannot parse into an AST does not need to be run as the thrown exception will be the same as the parse error.

We used the Python built-in \texttt{exec} function, isolating the global and local execution scopes.
To mitigate the impact of executing arbitrary source code, we ran the snippets in an isolated Linux kernel virtual machine (KVM) with no access to network.
Each snippet invocation was further containerized within its own Docker image running Python 3.6 using Debian Jessie.
We capped the maximum execution time to four seconds to account for long running and non-terminating programs.
All encountered exceptions and their corresponding stack traces are stored in our dataset for further analysis.

\section{Results}
We successfully parsed and analyzed 82,469,989 code snippets from the SOTorrent dataset.
Of these snippets, 7,260,631 contained a \texttt{python} matching tag.
A subset of 3,774,225 snippets parsed correctly.
Of the parseable snippets, 1,550,090 had prior versions, 62,965 of which contained a parse error.
A summary of the filtering steps can be found in Tab.~\ref{tab:filtering_sotorrent}.
With this dataset, we answer the four research questions.

\begin{table}[htbp]
    \caption{Filtered Dataset Entity Counts}
    \centering
    \tabcolsep=0.11cm
    \begin{tabular}{|c|c|}
    \hline
    \textbf{Filter Metric} & \textbf{Entity Block Count} \\
    \hline
    All code snippets & 82,469,989 \\
    Block matched \texttt{python} tag & 7,260,631 \\
    Content AST parseable & 3,774,225 \\
    Prior version exists & 1,550,090 \\
    Prior version AST error & 62,965 \\
    \hline
    \end{tabular}
    \label{tab:filtering_sotorrent}
\end{table}

\subsection*{RQ0. How accurate is our approach for extracting pairs of syntax errors and their fixes?}
We wanted to determine how accurate our methodology is for extracting true syntax errors and fixes.
Three co-authors together inspected 100 random samples of syntax errors and corrections from the extracted dataset of 62,965 pairs.
Only 2 observed cases were not entirely syntax error revisions, as they included large amounts of new code.
The 95\% confidence interval for our methodology is 0-5\% erroneous cases.

While our methodology is accurate, future research can take care of these erroneous cases (e.g., by leveraging the code snippets version similarity provided by SOTorrent).

\begin{tcolorbox}[sharp corners,top=1mm,bottom=1mm]
Stack Overflow is a good source for methodologically generating pairs of \emph{natural} syntax errors and fixes.
\end{tcolorbox}

\subsection*{RQ1. What are Python parse errors and corresponding runtime properties of their corrections in Stack Overflow?}
Stack Overflow posts are written and parsed using a subset of Markdown.
Code blocks are defined by using four space indentations, creating an indented code block, or by surrounding the text using three backtick characters (\texttt{`}) to create a fenced code block.
In addition, the Stack Overflow text editor does not provide code assistance features, such as syntax highlighting and code completion.

Of the 62,965 AST parse errors, there were 35,564 (56.5\%) syntax errors, 27,075 (43.0\%) indentation errors, and 326 (0.5\%) tab errors.
Over 41\% of all AST parse errors were \texttt{SyntaxError: invalid syntax}.
A detailed summary of the error types and messages are listed in Tab.~\ref{tab:parse_dominant_errors}.

Over a third of all corrected snippets still throw a Name error.
Roughly one third, or 21,332 corrected snippets ran in a Python interpreter with no errors.
Only 0.4\% of all evaluated snippets triggered an execution timeout by running for longer than four seconds.
The results from running all AST fixes in a Python interpreter are summarized in Tab.~\ref{tab:runtime_dominant_errors}.



\begin{tcolorbox}[sharp corners,top=1mm,bottom=1mm]
We extracted 62,965 parse errors and corrections from over 82 million code snippets.
Of the corrections, 21,332 ran in a Python interpreter without error.
\end{tcolorbox}


\begin{table}[htbp]
    \centering
    \caption{Top Python3.6 ast.parse Errors}
    \tabcolsep=0.11cm
    \begin{tabular}{|c|c|c|c|}
    \hline
    \textbf{Error} & \textbf{Message} & \textbf{Count} & \textbf{\%} \\
    \hline
    Syntax & invalid syntax & 26336 & 41.83 \\ \hline
    Indentation & expected an indented block & 14535 & 23.08 \\ \hline
    Indentation & unexpected indent & 11030 & 17.52 \\ \hline
    Syntax & Missing parentheses in call to 'print' & 3063 & 4.86 \\ \hline
    Syntax & unexpected EOF while parsing & 2732 & 4.34 \\ \hline
    Syntax & EOL while scanning string literal & 2045 & 3.25 \\ \hline
    \multirow{2}{*}{Indentation} & unindent does not match & \multirow{2}{*}{1468} & \multirow{2}{*}{2.33} \\
    & any outer indentation level & & \\ \hline
    \multirow{2}{*}{Tab} & inconsistent use of tabs & \multirow{2}{*}{326} & \multirow{2}{*}{0.52} \\
    & and spaces in indentation & & \\ \hline
    Syntax & invalid character in identifier & 297 & 0.47 \\ \hline
    \multirow{2}{*}{Syntax} & unexpected character after & \multirow{2}{*}{183} & \multirow{2}{*}{0.29} \\
    & line continuation character & & \\ \hline
    Syntax & invalid token & 139 & 0.22 \\ \hline
    \multirow{2}{*}{Syntax} & positional argument follows & \multirow{2}{*}{126} & \multirow{2}{*}{0.20} \\
    & keyword argument & & \\ \hline
    \multirow{2}{*}{Syntax} & EOF while scanning & \multirow{2}{*}{102} & \multirow{2}{*}{0.16} \\
    & triple-quoted string literal & & \\ \hline
    \multirow{2}{*}{Syntax} & (unicode error) 'unicodeescape' & \multirow{2}{*}{97} & \multirow{2}{*}{0.15} \\
    & codec can't decode bytes\dots & & \\ \hline
    Syntax & can't assign to function call & 94 & 0.15 \\ \hline
    Syntax & can't assign to operator & 63 & 0.10 \\ \hline
    Syntax & illegal target for annotation & 59 & 0.09 \\ \hline
    Syntax & keyword can't be an expression & 54 & 0.09 \\ \hline
    \multirow{2}{*}{Syntax} & Generator expression must be & \multirow{2}{*}{47} & \multirow{2}{*}{0.07} \\
    & parenthesized if not sole argument & & \\ \hline
    Indentation & unexpected unindent & 42 & 0.07 \\ \hline
    Syntax & \texttt{other}\dots & 127 & 0.20 \\ \hline
	\multicolumn{2}{|c|}{\textbf{Total}} & 62965 & 100 \\ \hline
    \end{tabular}
    \label{tab:parse_dominant_errors}
\end{table}

\begin{table}[htbp]
    \centering
    \caption{Top Python3.6 Corrected Code Runtime Errors}
    \begin{tabular}{|c|c|c|}
        \hline
        \textbf{Error} & \textbf{Count} & \textbf{\%} \\
        \hline
        NameError & 24270 & 38.55 \\ \hline
        \texttt{No Error} & 21332 & 33.88 \\ \hline
        ModuleNotFoundError & 9927 & 15.77 \\ \hline
        EOFError & 2223 & 3.53 \\ \hline
        FileNotFoundError & 2156 & 3.42 \\ \hline
        SyntaxError & 1013 & 1.61 \\ \hline
        TypeError & 567 & 0.90 \\ \hline
        TclError & 348 & 0.55 \\ \hline
        \texttt{Execution Timeout} & 251 & 0.40 \\ \hline
        AttributeError & 244 & 0.39 \\ \hline
        ImportError & 116 & 0.18 \\ \hline
        \texttt{other\dots} & 518 & 0.82 \\ \hline
        \textbf{Total} & 62965 & 100 \\ \hline
    \end{tabular}
    \label{tab:runtime_dominant_errors}
\end{table}

\subsection*{RQ2. Are Python errors in Stack Overflow similar to errors made by student programmers?}
We compared the distribution of our errors with the distribution of runtime errors as presented in two prior works.
If the distribution of errors are similar to student programmer errors then student errors can be viewed as representative.
If they are not similar then this provide evidence to arguments regarding how representative student errors are considering their error distribution is different.

\citeauthor{kelley2018system} collected the distribution of Python errors submitted by MIT programming students to an online tutor as part of an introductory Python programming course~\cite{kelley2018system}.
The collected student error distributions were TypeError: $28\%$; AttributeError: $22\%$; NameError: $19\%$; SyntaxError: $12\%$; IndexError: $6\%$ and other: $13\%$.
We performed a Pearson's Chi-squared Test for given probabilities, comparing our combined errors with the MIT student error distributions.
We found that the error distribution in Stack Overflow was not statistically similar to the errors made by MIT student programmers ($\alpha=0.01$, $X^2=239031$, $df=5$, $p$-value $<2.2\mathrm{e}{-16}$).

\citeauthor{Pritchard:2015:FDE:2846680.2846681} collected the student error distribution from interactive tutorials within CS Circles~\cite{Pritchard:2015:FDE:2846680.2846681}.
The paper reported distribution was SyntaxError: $48.14\%$; NameError: $15.11\%$; EOFError: $11.82\%$; IndentationError: $3.23\%$; and other: $21.7\%$.
We performed a Chi-squared test for the probabilities given in CS Circles student Python error distributions and found that errors in Stack Overflow were not similarly distributed ($\alpha=0.01$, $X^2=154689$, $df=4$, $p$-value $<2.2\mathrm{e}{-16}$).

The distribution of errors found in Stack Overflow extracted source code snippets did not match the distribution of errors prior literature has found in student and novice developers.
This empirically suggests that creating a corpus of syntax errors strictly from novice developers is insufficient to capture the full spectrum of errors made by the developer community as a whole.
Additional methods of contributing to existing syntax corpus datasets are therefore necessary.
We therefore emphasize the value of our corpus of data, as it provides a traceable distribution of \emph{natural} programming errors and corrections.

\begin{tcolorbox}[sharp corners,top=1mm,bottom=1mm]
The errors posted on Stack Overflow are not similar to the errors made by student and novice programmers.
\end{tcolorbox}

\subsection*{RQ3. Are Python errors in Stack Overflow similar to generated errors created by mutating valid code?}

\begin{figure}[htbp]
    \centerline{\includegraphics[width=3.5in]{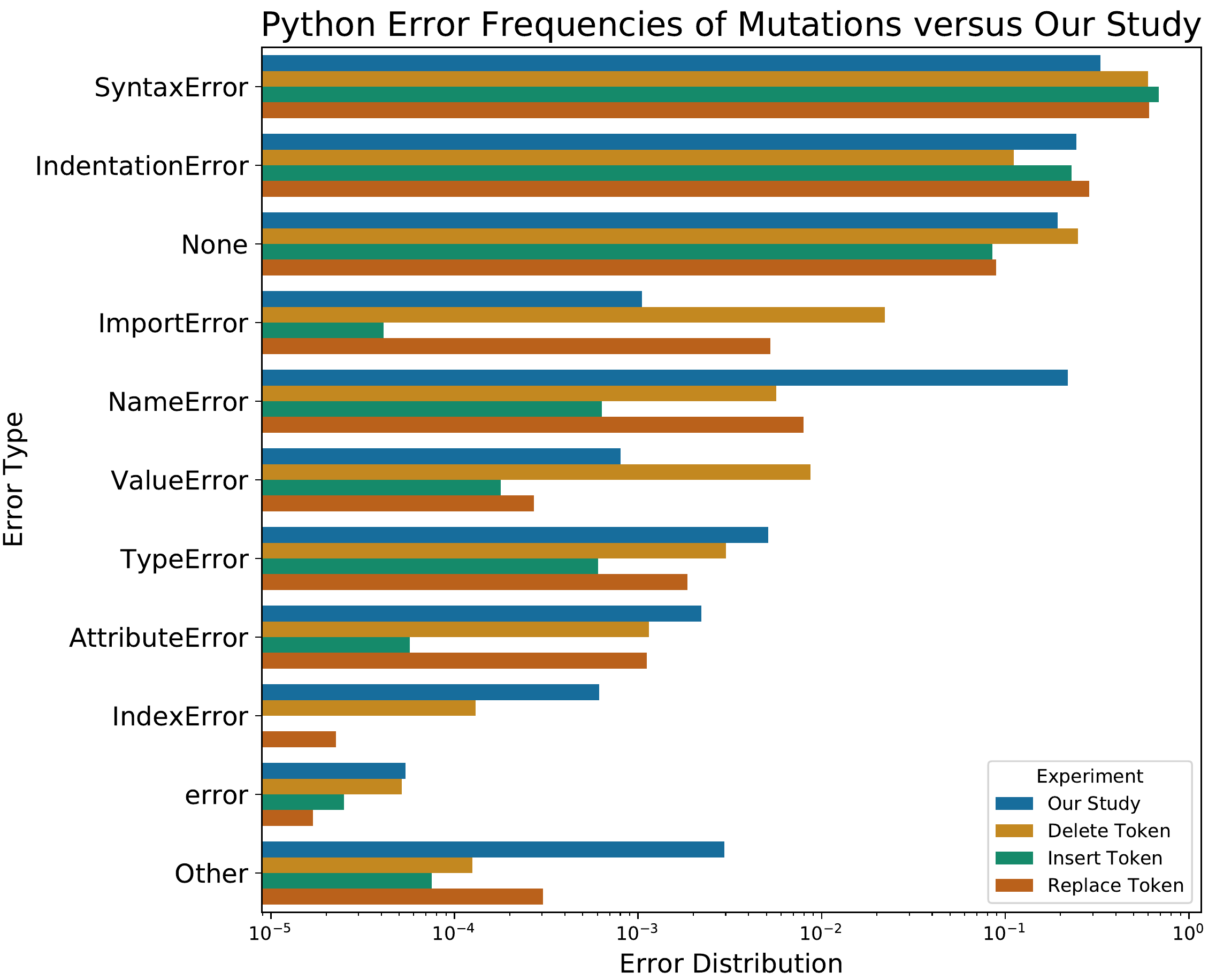}}
    \caption{Comparison of error distributions in our study vs random mutations}
    \label{fig:exception distrbution}
\end{figure}

We evaluated our errors against a distribution of syntax errors generated by random source code mutations, summarized in Fig.~\ref{fig:exception distrbution}.
\citeauthor{10.7287/peerj.preprints.1132v1} provided frequencies of Python exceptions caused by mutations of a randomly chosen token within a sequence of valid \texttt{Python} source code~\cite{10.7287/peerj.preprints.1132v1}.
We performed a Pearson's Chi-squared Test for the distributions of Token deletions, insertions, and replacements.

\begin{itemize}[]
    \item Delete; $X^2 = 933962$, $df = 10$, $p$-value $<2.2\mathrm{e}{-16}$
    \item Insert; $X^2 = 8424235$, $df = 10$, $p$-value $<2.2\mathrm{e}{-16}$
    \item Replace; $X^2 = 652478$, $df = 10$, $p$-value $<2.2\mathrm{e}{-16}$
\end{itemize}

We reject the assumption that the Stack Overflow extracted code error distributions are similar to the mutated source code using any mutation type with $\alpha = 0.01$.
This suggests that random mutations alone are insufficient for creating an error corpus that compares to natural faults.
As the distributions of mutation-based syntax errors were also not similar to errors found in programming novices, additional tuning is necessary in order to emulate natural distributions of syntax errors.

\begin{tcolorbox}[sharp corners,top=1mm,bottom=1mm]
The errors posted on Stack Overflow are not similar to generated errors from randomly mutating source code.
\end{tcolorbox}

\section{Future Work}
The methodology and data provided in this paper enable additional opportunities of research for software engineering.
\begin{itemize}
    \item The current paper generates syntax errors and corrections only for Python3.6. A natural, incremental addition is to perform the same computation for all Python versions.
    \item Further languages can also be analyzed using our methodology. These languages will have their own static parsing and runtime execution specifications.
    \item Extra runtime error corrections can be done by installing required dependencies or initializing variables.
    \item We will investigate how this dataset enhances existing code completion and syntax error location models, as well as provide a new evaluation benchmark for these tasks.
    \item More insight on \textit{natural} error corrections is crucial. Fixing an improperly indented block of code involves multiple white-space changes across many lines but can be resolved by highlighting the lines of code in an editor and indenting the selection. Character/token changes alone do not fully encapsulate developer-code interaction.
\end{itemize}

\section{Threats to Validity}
We relied on user submitted tags to identify Python code.
Because snippets not matching the Python tag were not considered, unaccounted Python code in Stack Overflow may still exist.
Unfortunately, relying only on source code parseability generated many false positives.
One recurring issue was JavaScript object notation (JSON) had the same structure as Python dictionaries.
Without matching the Python tag, an additional 5.4 million false positive code snippets would require analysis.


Another concern is whether or not we can rely on methodology to accurately distinguish between true syntactically invalid Python code versus syntactically invalid arbitrary text.
Although we found no occurrences of blatantly invalid source code, we publicize our dataset so anyone can audit our results.

Our approach obtains pairs of snippets where one snippet is syntactically correct and the other is not.
We only look at each snippet previous version at one point in time rather than looking at the entire history.
This prevents the inclusion of code blocks that have evolved greatly over time. 

Another hurdle is to obtain runtime faults from parseable Python code snippets, it is necessary to run the source code in a Python interpreter.
Despite our attempts to eliminate variability of source code execution using KVMs and Docker, we acknowledge that more work is necessary to rigorously evaluate runtime faults of programs.
Improvements involve accommodating operating system specificity, file system case sensitivity, and underlying hardware limitations.

We argue that the distribution of errors extracted from Stack Overflow are representative of developers, but not representative of software projects as a whole.
We acknowledge this distinction as code snippets are a subset of working software.


\section*{Conclusion}
We provide a novel methodology for automatically extracting natural source code syntax errors and their fixes from Stack Overflow, which sets the groundwork for future software naturalness and future syntax error detection and correction research.
Syntax errors extracted from Stack Overflow do not match prior distributions found in novice code or randomly generated errors.
We hope our data will be used for training and evaluating code completion \& error detection models, analysis of common programming language pitfalls, and future source code naturalness research.

\bibliographystyle{IEEEtranN}
\bibliography{main} 

\end{document}